\documentclass[prb,twocolumn,superscriptaddress,showpacs,aps]{revtex4-1}
\usepackage{graphicx,amsmath,amssymb,amsfonts}
\usepackage{textcomp}
\usepackage{array}
\usepackage{multirow}
\begin{document}

\title{Neutron diffraction study on magnetic structures and transitions in Sr$_{2}$Cr$_{3}$As$_{2}$O$_{2}$}

\author{Juanjuan Liu}
\thanks{These authors contributed equally to this work.}
\affiliation{Department of Physics, Renmin University of China, Beijing
	100872, China}
%juanjuanliu@ruc.edu.cn

\author{Jinchen Wang}
\thanks{These authors contributed equally to this work.}
\affiliation{Department of Physics, Renmin University of China, Beijing
	100872, China}
%jcwang_phys@ruc.edu.cn

\author{Jieming Sheng}
\affiliation{Department of Physics, Renmin University of China, Beijing
	100872, China}
\affiliation{Neutron Scattering Division, Oak Ridge National Laboratory, Oak Ridge, Tennessee 37831, USA}
%shengjm2011@gmail.com

\author{Feng Ye}
\affiliation{Neutron Scattering Division, Oak Ridge National Laboratory, Oak Ridge, Tennessee 37831, USA}
%yef1@ornl.gov

\author{Keith M. Taddei}
\affiliation{Neutron Scattering Division, Oak Ridge National Laboratory, Oak Ridge, Tennessee 37831, USA}
%taddeikm@ornl.gov

\author{J. A. Fernandez-Baca}
\affiliation{Neutron Scattering Division, Oak Ridge National Laboratory, Oak Ridge, Tennessee 37831, USA}
%fernandezbja@ornl.gov

\author{Wei Luo}
\affiliation{Key Laboratory of Neutron Physics, Institute of Nuclear Physics and Chemistry, China Academy of Engineering Physics, Mianyang 621999, China}
\affiliation{Department of Physics, Renmin University of China, Beijing
	100872, China}
%retornado@163.com

\author{Guang-Ai Sun}
\affiliation{Key Laboratory of Neutron Physics, Institute of Nuclear Physics and Chemistry, China Academy of Engineering Physics, Mianyang 621999, China}
%guangaisun_80@163.com

\author{Zhi-Cheng Wang}
\affiliation{Department of Physics, Zhejiang University, Hangzhou 310027, China}
%email: zcwang@zju.edu.cn 

\author{Hao Jiang}
\affiliation{School of Physics and Optoelectronics, Xiangtan University, Xiangtan 411105, China}
%email: yaal@163.com 

\author{Guang-Han Cao}
\affiliation{Department of Physics, Zhejiang University, Hangzhou 310027, China}
%email: ghcao@zju.edu.cn 

\author{Wei Bao}
\email{wbao@ruc.edu.cn}
\affiliation{Department of Physics, Renmin University of China, Beijing
	100872, China}

\date{\today}

\begin{abstract}
Sr$_{2}$Cr$_{3}$As$_{2}$O$_{2}$ is composed of alternating square-lattice CrO$_2$ and Cr$_2$As$_2$ stacking layers, where CrO$_2$ is isostructural to the CuO$_2$ building-block of cuprate high-$T_c$ superconductors and Cr$_2$As$_2$ to Fe$_2$As$_2$ of Fe-based superconductors. Current interest in this material is raised by theoretic prediction of possible superconductivity.
In this neutron powder diffraction study, we discovered that magnetic moments of Cr(II) ions in the Cr$_2$As$_2$ sublattice develop a C-type antiferromagnetic structure below 590 K, and the moments of Cr(I) in the CrO$_2$ sublattice form the La$_2$CuO$_4$-like antiferromagnetic order below 291 K. The staggered magnetic moment $2.19(4) \mu_{B}$/Cr(II) in the more itinerant Cr$_2$As$_2$ layer is smaller than $3.10(6) \mu_{B}$/Cr(I) in the more localized CrO$_2$ layer. Different from previous expectation, a spin-flop transition of the Cr(II) magnetic order observed at 291 K indicates a strong coupling between the CrO$_2$ and Cr$_2$As$_2$ magnetic subsystems.
\end{abstract}

%\pacs{ 74.10.+v, 75.25.-j, 75.50.Ee, 61.05.F-}

\maketitle

\section{INTRODUCTION}
It is well known that the square-lattice CuO$_2$ layer is the essential building block of cuprate superconductors \cite{Bednorz1986}. The antiferromagnetic order in the CuO$_2$ layer is suppressed by charge doping which gives rise to high-temperature superconductivity \cite{Lee2006}. For the more recently discovered iron pnictide superconductors, the common structural ingredient is the Fe$_2$As$_2$ layer \cite{LaOFeAs.2008}, and the interplay between antiferromagnetism in the layer and superconductivity has attracted intense research interest \cite{rev2009H,rev2010l,rev2011t,CPB2013}. 
The fact that unconventional superconductivity with high transition temperature ($T_c$) tends to occur in materials containing the CuO$_2$
or Fe$_2$As$_2$ structural ingredient has inspired investigation on materials of the same structural building blocks.
New type of unconventional superconductivity was discovered in Sr$_2$RuO$_4$ \cite{214RMP} and the stripe physics, a correlated charge and antiferromagnetic order phenomenon, was uncovered in Sr$_2$NiO$_4$ \cite{Ni214_Sr2}, when Cu in the cuprate Sr$_2$CuO$_4$ was replaced by Ru or Ni. Superconductivity has also been discovered in doped CoO$_2$ layered compounds \cite{H2OSC}.
Similarly, isostructural materials of the iron pnictide superconductors have also been the focus of recent research.
It has been proposed in theoretic
studies that superconductivity could be induced by electron doping LaCrAsO \cite{PhysRevB.95.144507.2017, PhysRevB.95.075115} and BaCr$_{2}$As$_{2}$ \cite{PhysRevB.95.205118}, similar to the `1111' and `122'-type Fe-based superconductors \cite{LaOFeAs.2008,A054630}.  
Experimental studies have been carried out so far on the parent compounds LaCrAsO \cite{LnCrAsO.2013}, BaCr$_{2}$As$_{2}$ \cite{PhysRevB.79.094429.2009}, SrCr$_{2}$As$_{2}$ \cite{PhysRevB.96.014411.2017} and EuCr$_{2}$As$_{2}$ \cite{PhysRevB.94.094411.2016}, and these materials are found to be metallic, and to develop a G-type antiferromagnetic order with the N\'{e}el temperature as high as 600 K.

An interesting family of oxipnictides $A_2$Mn$_3Pn_2$O$_2$ ($A$ = Sr, Ba and $Pn$ = P, As, Sb, Bi) \cite{Brechtel1979,Stetson1991, Brock1994, Brock1996, Brock_magstr1996, Ozawa1998, Matsushita20001424,Ozawa2001,PhysRevB.81.224513.2010,Eguchi2013} is formed of alternating CuO$_2$-type and Fe$_2$As$_2$-type layers, containing the key structural ingredients of both the cuprate and Fe-based superconductors. The two different types of Mn square lattices of MnO$_2$ and Mn$_2$As$_2$ stacking along the $c$ axis with the tetragonal space group $I4/mmm$ (No.~139). The isostructural `2322' oxysulfides also exist \cite{Zhu1997, ZHU1997319, Otzschi1999,Gal2006,Smura2011,Jin2012}.
Although $A_2$Mn$_3Pn_2$O$_2$ are antiferromagnetic insulators, it was suggested in theoretic works that superconductivity can be realized in the `2322' family with a proper choice of the transition metal to substitute Mn \cite{Volkova2008,Ozawa2008,XDai2016}.
Motivated by the theoretic suggestion, the chromium based `2322' material Sr$_{2}$Cr$_{3}$As$_{2}$O$_{2}$ was recently synthesized \cite{PhysRevB.92.205107.2015}, containing alternating CrO$_2$ and Cr$_2$As$_2$ layers along the $c$ axis, see Fig.~\ref{str} (a). The electrical resistivity shows metallic response \cite{PhysRevB.92.205107.2015}, as in the recently investigated `1111' and `122'-type chromium pnictides \cite{LnCrAsO.2013,PhysRevB.79.094429.2009,PhysRevB.96.014411.2017,PhysRevB.94.094411.2016}.

In $A_2$Mn$_3Pn_2$O$_2$, the Mn$_2$Pn$_2$ layer tends to form a G-type antiferromagnetic order around 300 K \cite{Brock_magstr1996}.
The magnetic couplings in the MnO$_2$ layer are significantly weaker.
While the MnO$_2$ layer in Sr$_2$Mn$_3$Sb$_2$O$_2$ forms long-range magnetic order at 65 K \cite{Brock_magstr1996}, it shows only short-range 
magnetic order with a spin freezing behavior below $\sim$50 K in compounds such as Sr$_2$Mn$_3$As$_2$O$_2$ \cite{Brock_magstr1996}, Sr$_2$Zn$_2$As$_2$MnO$_2$ \cite{Ozawa2001} and Ba$_2$Zn$_2$As$_2$MnO$_2$ \cite{PhysRevB.81.224513.2010}.

For Sr$_{2}$Cr$_{3}$As$_{2}$O$_{2}$, anomalies in electric resistivity, magnetic susceptibility and heat capacity were observed below the room temperature, indicating magnetic transitions in line with the prediction of magnetically ordered ground state in both the CrO$_2$ and Cr$_2$As$_2$ layers from first principle calculation \cite{PhysRevB.92.205107.2015}.
However, the nature of magnetic transitions at the bulk measurement anomalies remains unclear. 
Therefore we have conducted a neutron powder diffraction study on Sr$_{2}$Cr$_{3}$As$_{2}$O$_{2}$ to investigate
magnetic structures in the two Cr sublattices and the corresponding phase transitions.

We found that the magnetic moments of the Cr(II) ions in the Cr$_2$As$_2$ sublattice develop a C-type antiferromagnetic order at a temperature as high as 590 K with the $c$-axis as the magnetic easy axis, see Fig.~\ref{str} (c). 
The in-plane magnetic arrangement of Sr$_{2}$Cr$_{3}$As$_{2}$O$_{2}$ is the same as in the G-type antiferromagnetic order of the `1111' and `122'-type chromium pnictides \cite{LnCrAsO.2013,PhysRevB.79.094429.2009,PhysRevB.96.014411.2017,PhysRevB.94.094411.2016}, however, the inter-plane Cr(II) ions
are ferromagnetically aligned in contrast to the case in the G-type order in those chromium pnictides.
Below 291 K, the Cr(I) ions in the CrO$_2$ sublattice forms a K$_2$NiF$_4$-type \cite{PhysRevB.1.2211} long-range antiferromagnetic order with moments along the $c$ direction, see Fig.~\ref{str} (b). This is the same antiferromagnetic order discovered in the cuprate La$_2$CuO$_4$ \cite{la3dv}.
When the Cr(I) sublattice orders at 291 K, 
a spin-flop transition occurs on the Cr(II) sublattice with the moment direction flips from the $c$ axis to the $ab$ plane.
The observed magnetic structure is different from the prediction from the first-principle calculations \cite{PhysRevB.92.205107.2015}.

\section{EXPERIMENTAL METHODS}
The polycrystalline sample of Sr$_{2}$Cr$_{3}$As$_{2}$O$_{2}$ was grown by solid-state reactions using SrO (99.5\%), Cr (99.9\%) and As (99.999\%) as starting materials \cite{PhysRevB.92.205107.2015}. 
The sample was initially examined using the high-resolution neutron powder diffractometer at the Key Laboratory of Neutron Physics, China Academy of Engineering Physics (CAEP).
Neutron powder diffraction measurements were carried out on the HB-2A neutron powder diffractometer \cite{HB2Ainstrument} at the High Flux Isotope Reactor (HFIR), Oak Ridge National Laboratory (ORNL). 
On HB-2A, neutron wavelength of $\lambda = 2.4101 \AA$ and $1.5392 \AA$ were selected by the vertically focused Ge(113) and Ge(115) monochromator, respectively. 
The beam was collimated by the 21$'$ Soller collimator before sample and 12$'$ before detectors. 
%The detector array composed of 44 $^{3}$He tubes was scanned in steps of 0.05\textdegree, covering 2$\theta$ range from 3\textdegree to 120\textdegree.
Approximate 5.74 g powder sample was sealed in a cylindrical Al container with Helium as exchange gas to ensure thermal equilibrium. It was loaded into a closed cycle refrigerator that regulates the temperature from 4 K to 300 K.
High temperature neutron data were taken on the HB-1A triple-axis spectrometer at HFIR with fixed incident neutron energy 14.64 meV (wavelength $ 2.364 \AA$) produced by the double pyrolytic graphite monochromator system.
The temperature was regulated by high temperature Displex from 30 K to 700 K.

Besides main phase of Sr$_{2}$Cr$_{3}$As$_{2}$O$_{2}$ and minor impurity phase, notable Al Bragg peaks from the sample environment were detected.
Using FullProf Suite package \cite{RODRIGUEZCARVAJAL199355}, the nuclear and magnetic structures of the main phase were refined using the Rietveld method \cite{Rietveld1969} while Al peaks are accounted by Le Bail fitting \cite{LeBail1988}.
The Thompson-Cox-Hastings pseudo-Voigt function corrected by axial divergence asymmetry was used to model the peak profiles of all phases.
Due to the existence of Al and impurity peaks, the refinements of thermal factors against individual atoms give unreliable results, so the overall thermal factor for all atoms is used at each temperature.
BasIreps program of FullProf Suite is used for representation analysis to derive the possible magnetic structure modes.

\section{Results and Discussions}
 
\begin{figure}[!tbp]
	\centering
	\includegraphics[width=\columnwidth]{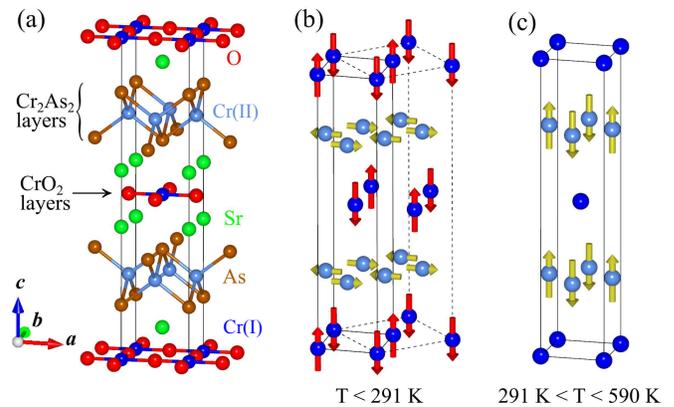}
	\caption{(color online) (a) Crystal structure of Sr$_{2}$Cr$_{3}$As$_{2}$O$_{2}$. The atoms are noted by different colors. Layers of CrO$_{2}$ squares and Cr$_{2}$As$_{2}$ tetrahedra are labeled.
		(b) The magnetic structure below 291 K when both CrO$_{2}$ layer and Cr$_{2}$As$_{2}$ layer order. (c) The magnetic structure above 291 K when only the Cr$_{2}$As$_{2}$ layer orders. The solid lines in (b) and (c) are the structural unit cell identical to the cell in (a), and the dash lines in (b) are the magnetic unit cell with $ \boldsymbol{a}_{mag} = \boldsymbol{a} + \boldsymbol{b} $, $ \boldsymbol{b}_{mag} = \boldsymbol{a} - \boldsymbol{b} $.
	}
	\label{str}
\end{figure}

\begin{figure}[!tbp]
	\centering
	\includegraphics[width=\columnwidth]{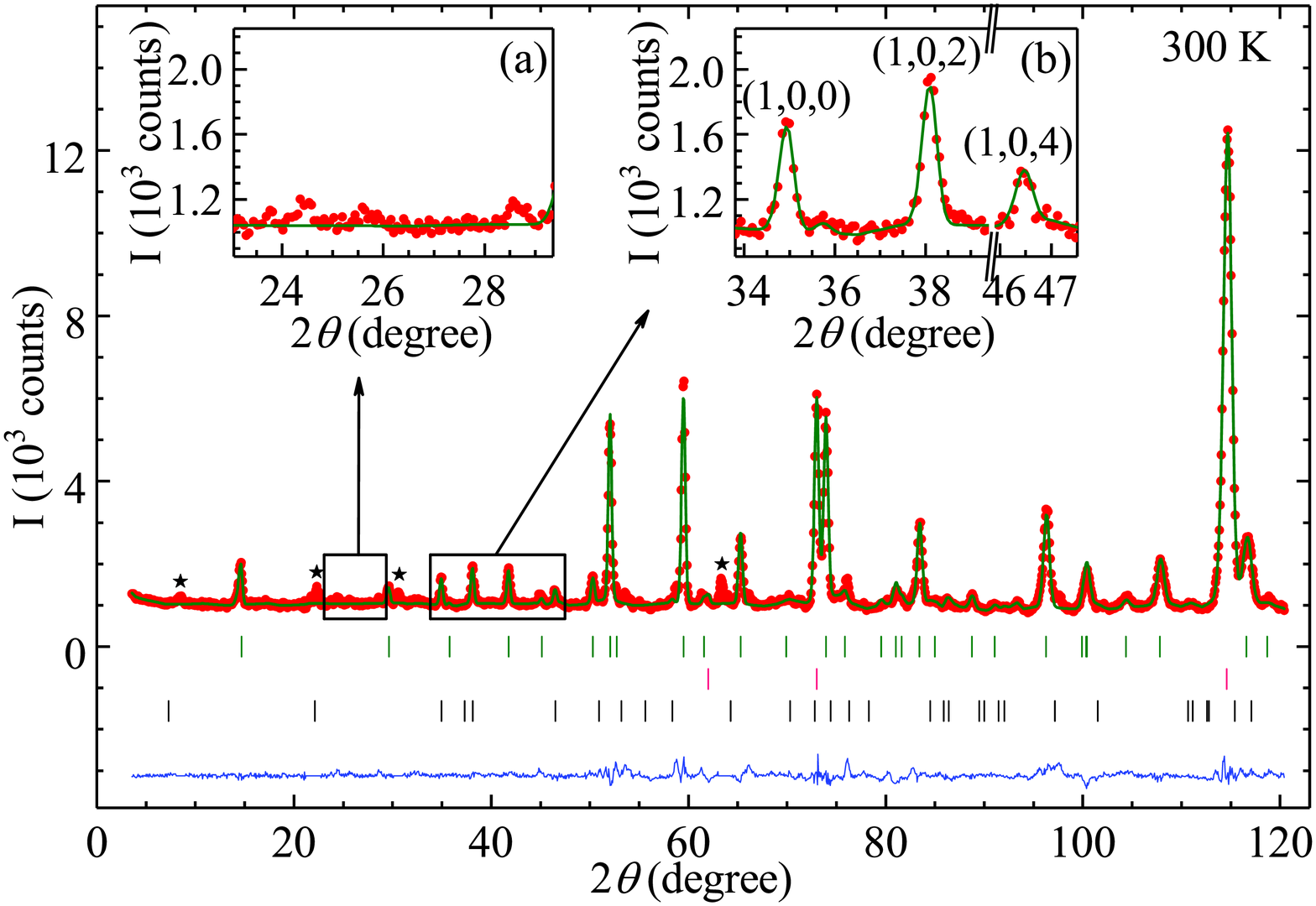}
	\caption{(color online) The neutron diffraction pattern of Sr$_{2}$Cr$_{3}$As$_{2}$O$_{2}$ at room temperature, with wavelength $\lambda = 2.4101 \AA{}$. 
		The observation, calculation and their difference are denoted in red circles, green and blue lines, respectively.
		The vertical bars in green, red and black mark the nuclear Bragg peaks positions of Sr$_{2}$Cr$_{3}$As$_{2}$O$_{2}$, Al and magnetic Bragg peaks positions from the Cr(II) sublattice. 
		The impurity peaks are marked with stars and are excluded during refinement.
		The inset (a) zooms in expected positions of magnetic peaks of wave-vector $(\frac{1}{2},\frac{1}{2},0)$ and inset (b) magnetic peaks of wave-vector $(1,0,0)$.
	}
	\label{rf300K}
\end{figure}

The neutron powder diffraction pattern measured at room temperature is shown in Fig.~\ref{rf300K}. 
At 300 K, two wavelengths were used, as the shorter wavelength of neutron probes the structure in a larger reciprocal-space range.
The refined parameters under different conditions are summarized in Table \ref{str_tb} and two wavelengths give essentially no different results.
Consistent with earlier reports \cite{PhysRevB.92.205107.2015}, Sr$_{2}$Cr$_{3}$As$_{2}$O$_{2}$ crystallizes in the tetragonal space group $I4/mmm $ (No.~139). 
The Cr(I) in the CrO$_{2}$ layer resides at the Wyckoff $2a$ site, forms body centered sublattice. 
The Cr(II) in the Cr$_{2}$As$_{2}$ layer resides at the Wyckoff $4d$ site, forms primitive sublattice.

Both the As-M-As bond angle (M denotes transition metal) and the As height are important parameters controlling the electronic states driven by the $p-d$ hybridization and exchange interactions.
In the iron-based superconductors, the As-Fe-As bond angle and the anion height were found in close relation to superconducting transition temperatures $T_c$. 
To achieve optimal $T_c$, it was suggested to have regular FeAs$_4$ tetragonal with the As-Fe-As bond angle close to 109.47\textdegree \cite{FeAsangle1111,qiu2008} or anion height close to 1.38 \AA \cite{mizuguchi2010}.
Here, the As-Cr(II)-As bond angle of Sr$_{2}$Cr$_{3}$As$_{2}$O$_{2}$ is around 105.8\textdegree and 111.3\textdegree, and the As height is around 1.51 $\AA$, with little temperature dependence from 300 K to 4 K as seen from Table \ref{str_tb}. 

\begin{table*}[!tbp]
	\caption{The Rietveld refinement results of Sr$_{2}$Cr$_{3}$As$_{2}$O$_{2}$ from the neutron powder diffraction data at various temperatures and wavelengths. 
		The atomic positions are Sr (0,0,$z_\text{Sr}$) in Wyckoff $4e$ site, Cr(I) (0,0,0) in $2a$ site, Cr(II) (0,0.5,0.25) in $4d$ site, As (0,0,$z_\text{As}$) in $4e$ site, O (0,0.5,0) in $4c$ site. 
		The two fold and four fold multiplicities of As-Cr(II)-As angles are denoted.
		The number in parentheses represents estimated standard deviation.
	}
	\label{str_tb} 
	\centering
	\begin{ruledtabular}   
		\begin{tabular}{cccccccc}  
			T (K) & 300 & 300 & 260 & 230 & 200 & 100 & 4 \\ 
			wavelength (\AA{}) & 1.5392  & 2.4101 & 2.4101 & 2.4101 & 2.4101 & 2.4101 & 2.4101 \\
			\hline 
			$a$ (\AA{}) & 4.00671(6) & 4.00574(5) & 4.00417(6) & 4.00306(6) & 4.00198(6) & 3.99941(6) & 3.99892(6)  \\  
			$c$ (\AA{}) & 18.8310(5) & 18.8261(5) & 18.8029(5) & 18.7862(5) & 18.7717(5) & 18.7303(5) & 18.7151(5)  \\ 
			$V$ (\AA{}$ ^{3} $) & 302.31(1) & 302.08(1) & 301.47(1) & 301.04(1) & 300.64(1) & 299.60(1) & 299.28(1)  \\  
			$z_\text{Sr}$ & 0.4125(2) & 0.4117(2) & 0.4120(2) & 0.4128(2) & 0.4123(2) & 0.4115(2) & 0.4113(2)  \\   
			$z_\text{As}$ & 0.8298(2) &  0.8308(2) & 0.8305(2) & 0.8304(2) & 0.8304(2) & 0.8309(2) & 0.8308(2)  \\ 
			B (\AA{}$ ^{2} $) & 0.77(2) & 0.60(4) & 0.47(4) & 0.42(4) & 0.34(4) & 0.30(4) & 0.18(4)   \\  
			As height (\AA{})  & 1.503(3)  & 1.521(4) & 1.514(4) & 1.510(4) & 1.510(4) & 1.515(4) & 1.512(4) \\
			As-Cr(II)-As angle (deg) $\times 2$  & 106.23(6) & 105.57(7)  &  105.82(7)  &  105.92(7)  &  105.95(7)  &  105.69(7)  &  105.80(7)    \\  
			As-Cr(II)-As angle (deg) $\times 4$ &111.1(1)  & 111.5(2)  &  111.3(2)  &  111.3(2)  &  111.3(2)  &  111.4(2)  &  111.3(2)    \\ 
			$m$ $(\mu_{B}/\text{Cr(II)})$ & 1.97(4) &  1.97(3) & 2.02(4) & 2.11(4) & 2.12(4) & 2.18(4) & 2.19(4)  \\
			$m$ $(\mu_{B}/\text{Cr(I)})$ & 0 &  0 & 2.46(7) & 2.69(7) & 2.78(6) & 3.07(6) & 3.10(6)  \\   
			R$ _{p}$ (\%) & 4.01 &  3.94 & 4.04 & 4.36 & 4.07 & 4.24 & 4.48  \\  
			R$ _{wp}$ (\%) & 5.43 &  5.55 & 6.04 & 6.48 & 6.44 & 6.33 & 6.53  \\      
		\end{tabular} 
	\end{ruledtabular}
\end{table*}

\begin{figure}[!tbp]
	\centering
	\includegraphics[width=.9\columnwidth]{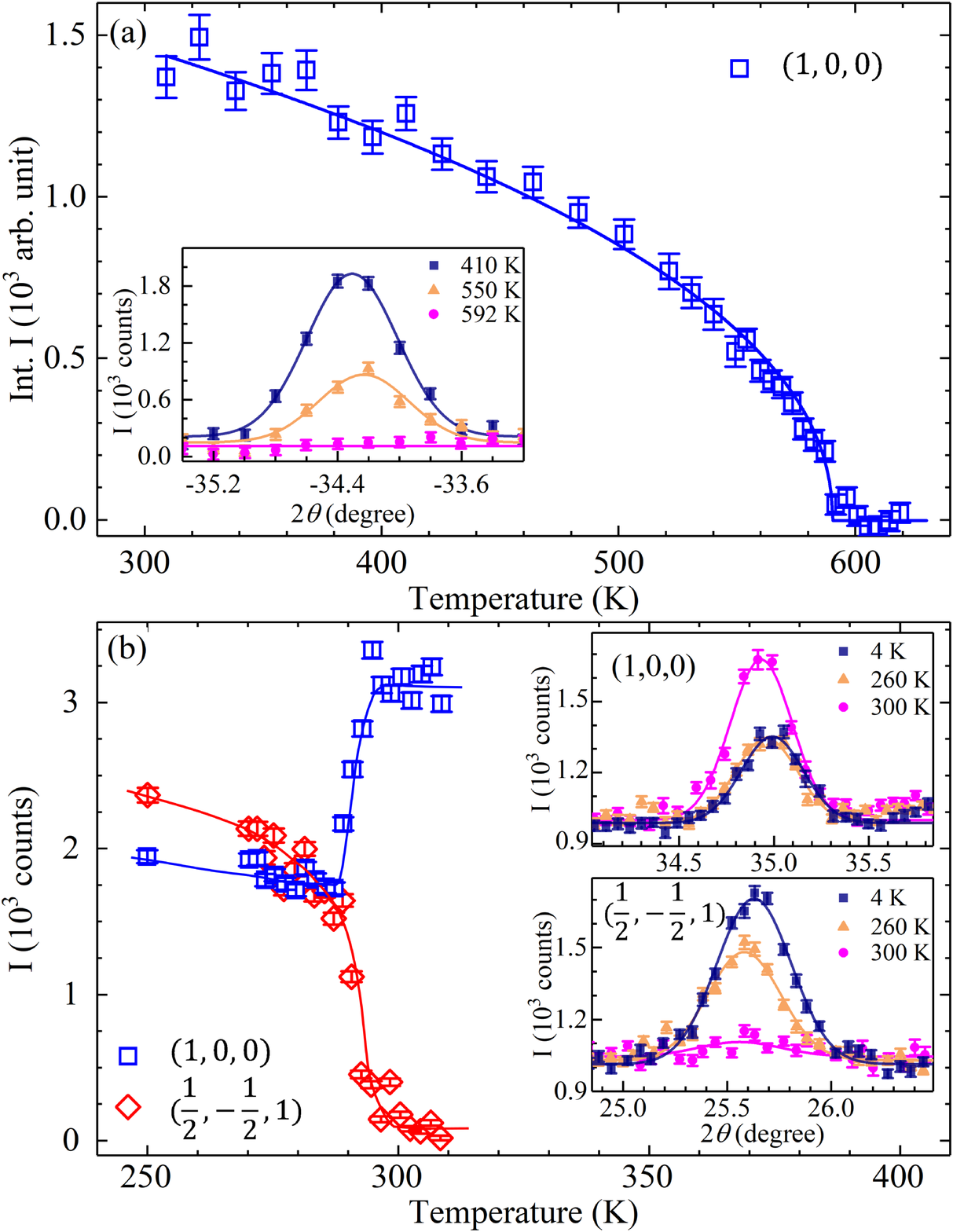}
	\caption{(color online) 
		(a) Temperature dependence of the integrated intensity of the magnetic Bragg peak (1,0,0). The solid line and onset temperature of $T_N = 590.3(7)$ K was obtained by fitting to the order parameter function $I/I_0 = (1-\frac{T}{T_N})^{2\beta} + b_0$.
		(b) Temperature dependence the background corrected peak intensity of the Bragg peaks (1,0,0) and $(\frac{1}{2}, -\frac{1}{2}, 1)$. The solid lines are guide to eyes.
		The insets are the peaks profiles for (1,0,0) and $(\frac{1}{2}, -\frac{1}{2}, 1)$ at selected temperatures.}
	\label{op}
\end{figure}

Three peaks indexed as (1,0,0), (1,0,2) and (1,0,4) that are forbidden by the crystal symmetry were observed at 300 K, as presented in the inset (b) of Fig.~\ref{rf300K}. 
These intensities are accounted by the C-type magnetic structure of the Cr(II) spins in the Cr$_2$As$_2$ layer shown in Fig.~\ref{str} (c).
It is antiferromagnetically (AFM) arranged for intra-layer nearest neighbor spins but ferromagnetically (FM) arranged for inter-layer spins, resulting in wave-vector (1,0,0).
The moment direction is along the $c$ axis, and the moment size is $1.97(4) \mu_{B}/\text{Cr(II)}$ at 300 K.
The magnetic structure persists up as high as $590.3(7)$ K, as demonstrated by the temperature dependence of (1,0,0) peak in Fig.~\ref{op} (a).

The high ordering temperature and the intra-layer AFM pattern established in the Cr$_2$As$_2$ layer of Sr$_{2}$Cr$_{3}$As$_{2}$O$_{2}$ resemble those in other chromium pnictides. 
In `1111'-type LaCrAsO \cite{LnCrAsO.2013}, and the `122'-type BaCr$_{2}$As$_{2}$, SrCr$_{2}$As$_{2}$ and EuCr$_{2}$As$_{2} \cite{PhysRevB.79.094429.2009,PhysRevB.96.014411.2017,PhysRevB.94.094411.2016}$, the same intra-layer spin arrangement and  ordering temperature around 600 K were found.
This reveals a robust AFM intra-layer coupling which does not change across compositions.
However, in Sr$_{2}$Cr$_{3}$As$_{2}$O$_{2}$ the inter-layer coupling becomes FM, in contrast to the `1111' and `122' chromium pnictides which have AFM inter-layer couplings in the G-type magnetic structure.

Below 291 K, new peaks appear that can be indexed by the wave-vector $\boldsymbol{k}' = (\frac{1}{2}, \frac{1}{2}, 0)$. 
Meanwhile, the $\boldsymbol{k}'' = (1,0,0)$ series of peaks change their intensity. 
This is demonstrated in Fig.~\ref{op} (b) by abrupt changes of magnetic peaks intensities on (1,0,0) and $(\frac{1}{2}, -\frac{1}{2}, 1)$ around 291 K.
As we will demonstrate, the arise of $(\frac{1}{2}, \frac{1}{2}, 0)$ series of peaks corresponds to the magnetic order established on the Cr(I) sublattice in the CrO$_2$ layer, and the intensity change on $(1,0,0)$ series of peaks corresponds to a sudden flip of Cr(II) moment in the Cr$_2$As$_2$ layer.
The transition is responsible for the susceptibility anomaly observed around 291 K\cite{PhysRevB.92.205107.2015}.
The first-order-like transitions in both the $(\frac{1}{2}, \frac{1}{2}, 0)$ and $(1,0,0)$ type magnetic Bragg peaks (Fig.~\ref{op} (b)) suggest strong interaction between the Cr(I) and Cr(II) ions.

\begin{figure}[!t]
	\centering
	\includegraphics[width=\columnwidth]{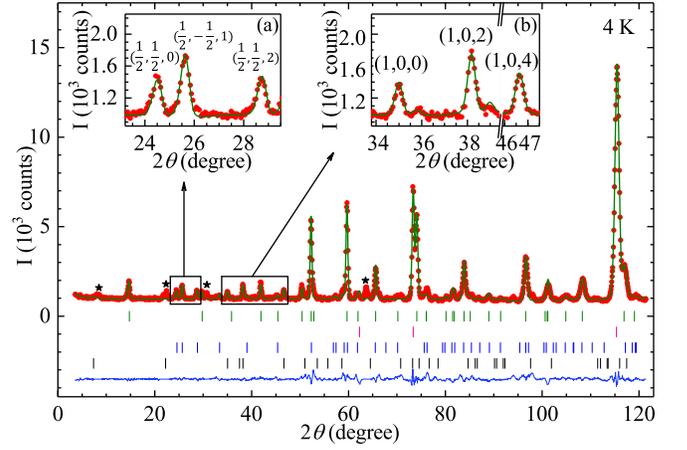}
	\caption{(color online) The observed and calculated neutron diffraction pattern of Sr$_{2}$Cr$_{3}$As$_{2}$O$_{2}$ at $T = 4$ K, with wavelength $\lambda = 2.4101 \AA{}$.
		The observation, calculation and their difference are denoted in red circles, green and blue lines, respectively.
		The vertical bars in green and red mark the nuclear Bragg peaks positions of Sr$_{2}$Cr$_{3}$As$_{2}$O$_{2}$ and Al, respectively.
		The vertical bars in blue and black mark the magnetic Bragg peaks positions from the Cr(I) and Cr(II) sublattice, respectively. 			
		The impurity peaks are marked with stars and are excluded during refinement.
		The inset (a) zooms in magnetic peaks of wave-vector $(\frac{1}{2},\frac{1}{2},0)$ and inset (b) magnetic peaks of wave-vector $(1,0,0)$. }
	\label{rf4K}
\end{figure}

Magnetic Bragg peaks below 291 K are fitted by the magnetic structure depicted in Fig.~\ref{str} (b), with the Cr(I) and Cr(II) moment sizes the only two refined parameters.
The magnetic order on the Cr(I) site of the CrO$_2$ layer is responsible for the $(\frac{1}{2}, \frac{1}{2}, 0)$ wave-vector.
The nearest neighbors of intra-layer Cr(I) spins are AFM aligned, and the moments point to $c$ axis. 
The magnetic structure of the body centered Cr(I) sublattice mimics that of the body centered K$_2$NiF$_4$ \cite{PhysRevB.1.2211}.
Interestingly, when the Cr(I) sublattice orders, the moments at the Cr(II) site in the Cr$_2$As$_2$ layer flip into the $ab$ plane. 
The magnetic structure where two sublattices have orthogonal spin orientations persists down to 4 K, the base temperature in our measurement.
The diffraction pattern together with the fits at 4 K are presented in Fig.~\ref{rf4K}. 
In principle, the specific in-plane moment direction of tetragonal structure cannot be resolved from powder diffraction data due to the powder-averaged intensity \cite{Shirane1959}. 
We therefore don't distinguish the in-plane moment direction of Cr(II), and the Cr(II) moment direction in Fig.~\ref{str} (b) is plotted for illustration purpose.
The refined moments on the Cr(I) and Cr(II) sublattices at each temperatures are summarized in Table \ref{str_tb}. 
From 300 K to 260 K, a quick increase of Cr(I) moment size is witnessed when the order is established. This is then followed by a slower increase rate and gradual saturation down to the base temperature. 
For the Cr(II) moments, despite an abrupt change of the direction, the ordered moment of Cr(II) shows very small change below 300 K.

\begin{table*}[!tbp]
	\caption{The basis vectors (BV) of decomposed irreducible representations (IR) of space group $I 4/mmm $ (No. 139) with wave-vector $(\frac{1}{2}, \frac{1}{2}, 0)$ and moments on $2a$ site (Cr(I) in CrO$_2$ layer), as well as wave-vector $(1, 0, 0)$ and moments on $4d$ site (Cr(II) in Cr$_2$As$_2$ layer). The corresponding magnetic structures are displayed. The solid lines depicted the crystallographic unit cell, while the dash line the magnetic cell. }
	\label{irrep_tb}
	\begin{tabular}{m{2cm}<{\centering}*{4}{m{3.8cm}<{\centering}}}
		\hline\hline
		\multirow{2}{*}{Irrep} & \multicolumn{4}{c}{Cr(I) at $ 2a $ site, $\boldsymbol{k}' = (\frac{1}{2}, \frac{1}{2}, 0)$}   \\ 
		\cline{2-5} &  $ \Gamma'_{3} $ & $ \Gamma'_{5} $ & $ \Gamma'_{7} $ &   \\
		\hline BV for site \  \  (0,0,0) &  $(0, 0, m_z)$  &  $(m_x, m_x, 0)$  &  $(m_x, -m_x, 0)$  &        \\
		\hline magnetic structure &  {\includegraphics[width=22mm]{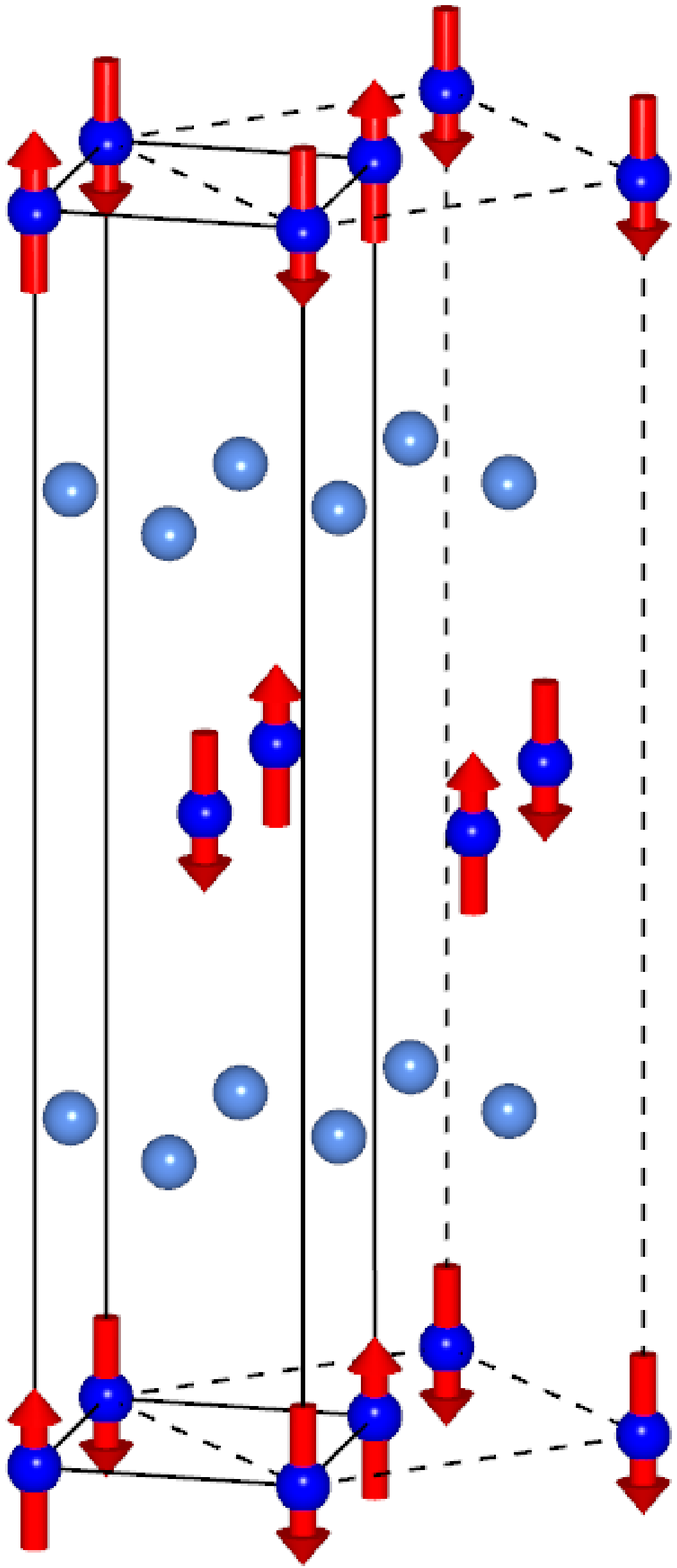}} &  {\includegraphics[width=22mm]{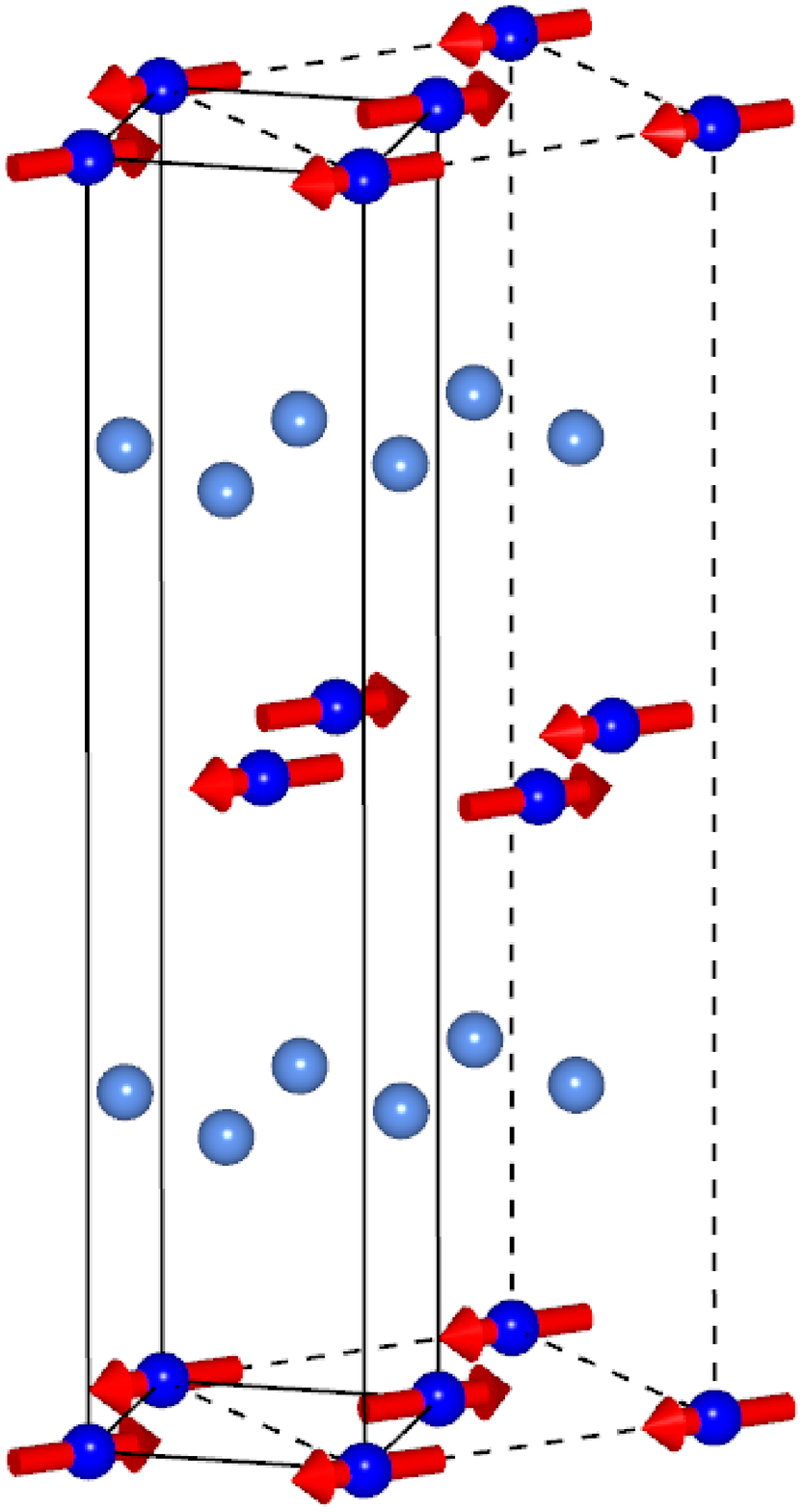}} &  {\includegraphics[width=22mm]{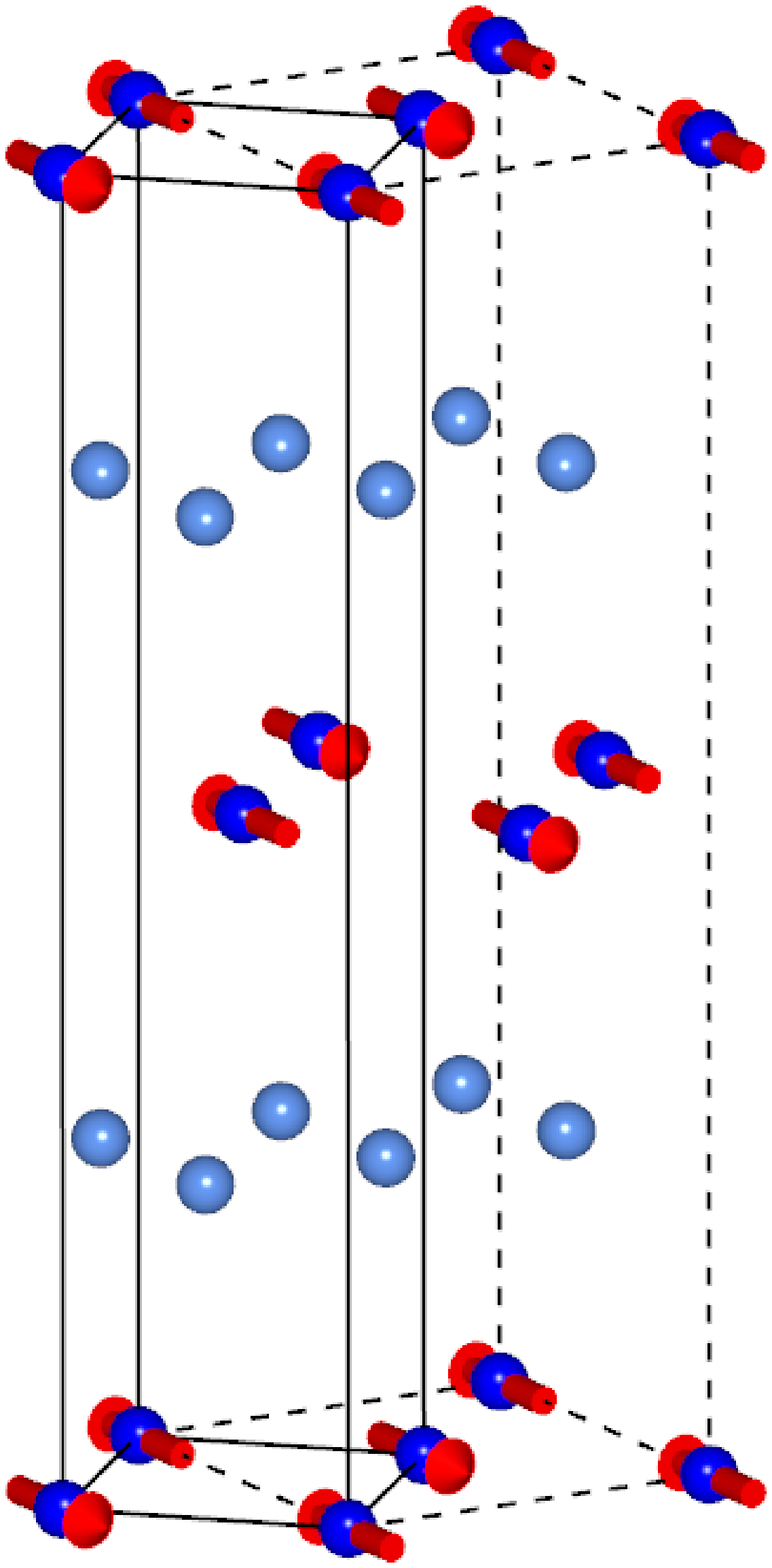}} &   \\ 
		\hline \multirow{2}{*}{Irrep}  & \multicolumn{4}{c}{Cr(II) at $ 4d $ site, $\boldsymbol{k}'' = (1, 0, 0)$} \\
		\cline{2-5}  & $ \Gamma''_{2} $ & $ \Gamma''_{5} $ & $ \Gamma''_{9} $ & $ \Gamma''_{10} $  \\
		\hline BV for site \  (0,0.5,0.25) &  $(0, 0, m_z)$  &  $(0, 0, m_z)$  &  $(m_x, 0, 0)$  $(0, m_y, 0)$  &  $(m_x, 0, 0)$  $(0, m_y, 0)$    \\
		BV for site \  (0,0.5,0.75) & $(0, 0, -m_z)$  &  $(0, 0, m_z)$  &  $(-m_x, 0, 0)$  $(0, -m_y, 0)$  &   $(m_x, 0, 0)$  $(0, m_y, 0)$    \\
		\hline magnetic structure mode  &  {\includegraphics[width=15mm]{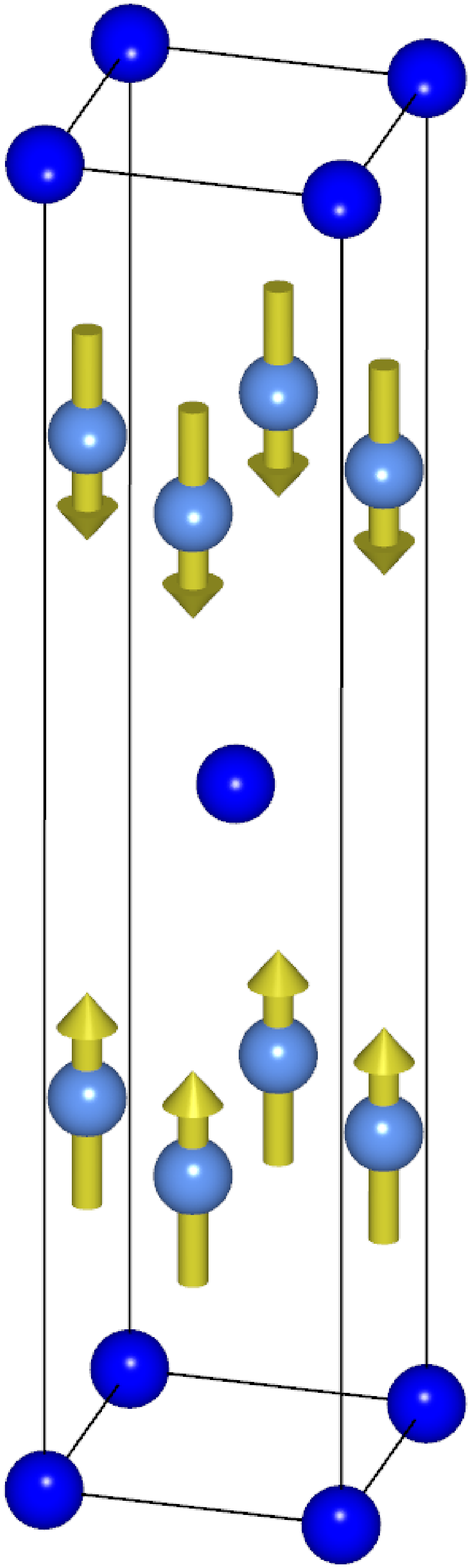}} &  {\includegraphics[width=15mm]{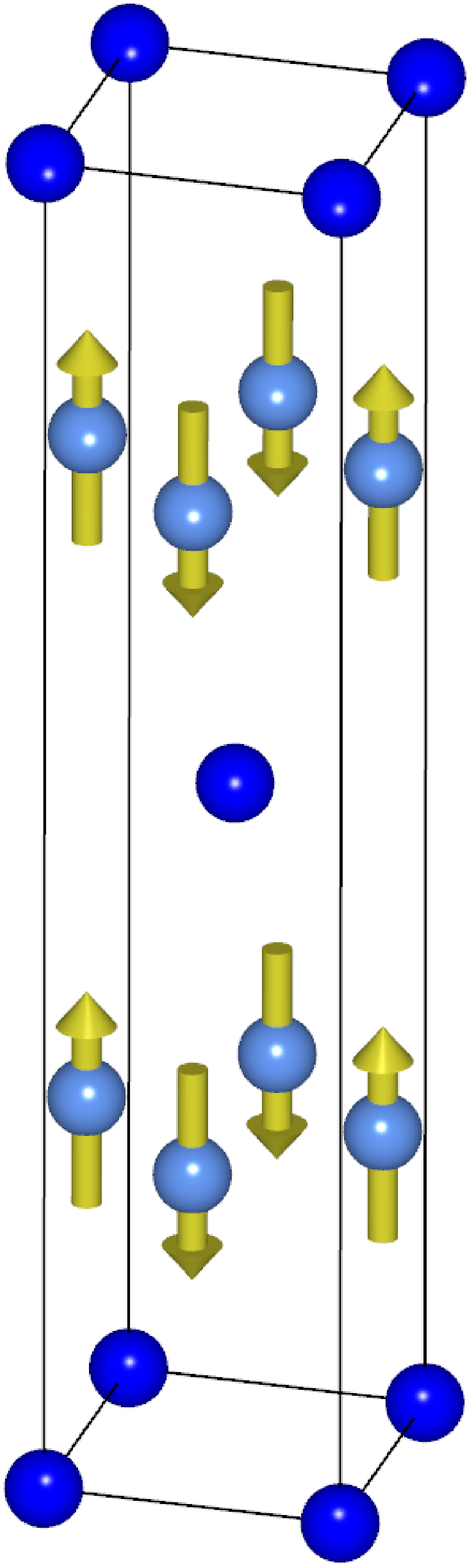}} &  {\includegraphics[width=15mm]{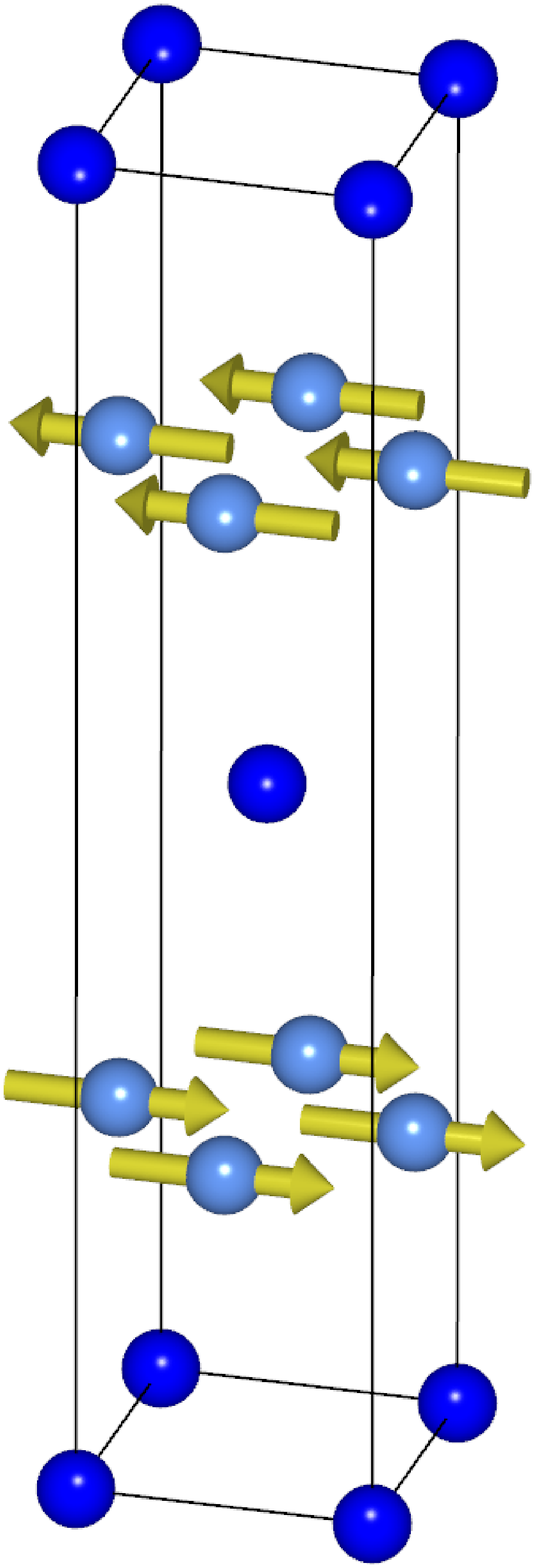}} &  {\includegraphics[width=15mm]{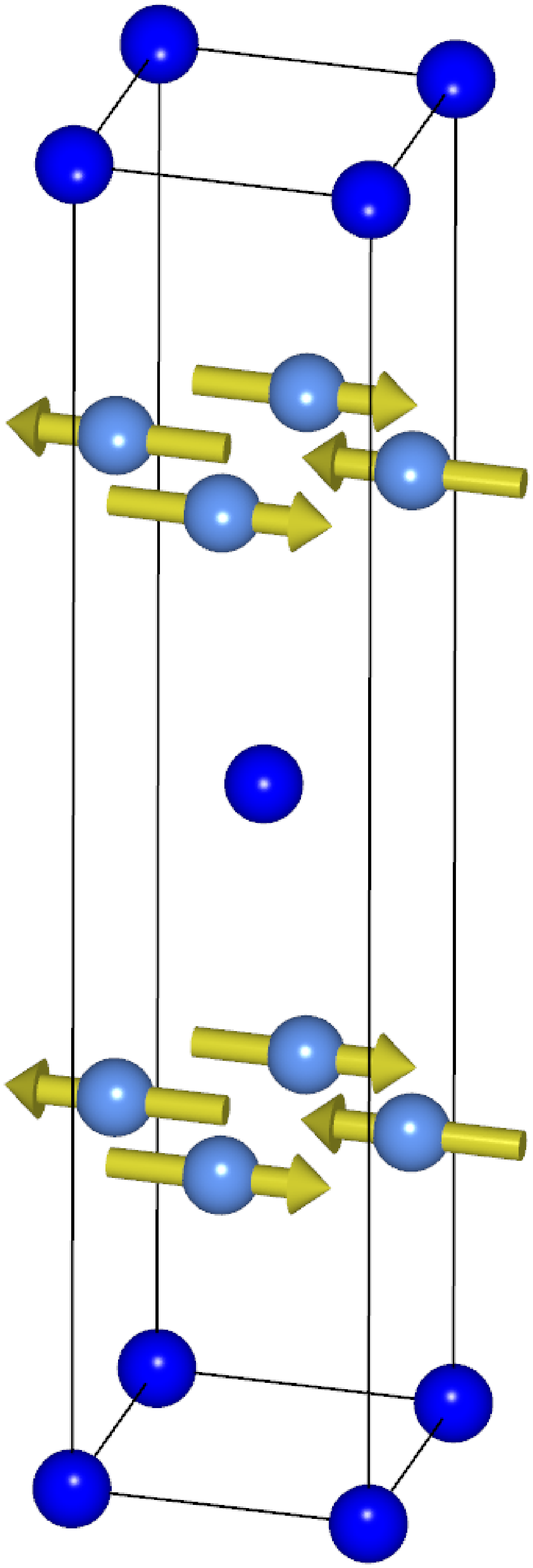}}  \\ 
		\hline\hline
	\end{tabular}
\end{table*}

The magnetic structures above and below 291 K are further confirmed by the representation analysis \cite{Bertaut.a05871}. 
The space group for the crystal structure is $I4/mmm$. 
For wave-vector $\boldsymbol{k}' = (\frac{1}{2}, \frac{1}{2}, 0)$, the little group $mmm$ that leaves the wave-vector unchanged has eight one-dimensional irreducible representations (IR) $\Gamma'_{1} \dots \Gamma'_{8}$. 
The linear spaces spanned by spin vectors on site Cr(I) and Cr(II) are three dimensional and six dimensional, respectively. 
Therefore, they can be decomposed to three and six one-dimensional IRs of group $mmm$
\begin{eqnarray*}
	\Gamma'_{\text{Cr(I)}} = 1\Gamma'_{3} + 1\Gamma'_{5} + 1\Gamma'_{7},
\end{eqnarray*}
and
\begin{eqnarray*}
	\Gamma'_{\text{Cr(II)}} =  1\Gamma'_{1} +  1\Gamma'_{2} + 1\Gamma'_{3} + 1\Gamma'_{4} + 1\Gamma'_{7} + 1\Gamma'_{8}.
\end{eqnarray*} 
Only the magnetic structure of $\Gamma'_{3}$ for Cr(I) spins fits the magnetic Bragg peaks of $(\frac{1}{2}, \frac{1}{2}, 0)$ series.
For wave-vector $\boldsymbol{k}'' = (1, 0, 0)$, the little group is the same as the crystal point group $4/mmm$. 
It has eight one-dimensional IRs $\Gamma''_{1} \dots \Gamma''_{8}$ and two two-dimensional IRs $\Gamma''_{9}$ and $\Gamma''_{10}$. 
The spin spaces on site Cr(I) and Cr(II) can be decomposed as 
\begin{eqnarray*}
	\Gamma''_{\text{Cr(I)}} = 1\Gamma''_{7} + 1\Gamma''_{10},
\end{eqnarray*}
and
\begin{eqnarray*}
	\Gamma''_{\text{Cr(II)}} =  1\Gamma''_{2} +  1\Gamma''_{5} + 1\Gamma''_{9} + 1\Gamma''_{10}.
\end{eqnarray*}
The magnetic Bragg peaks of $(1, 0, 0)$ series above 291 K comes from the $\Gamma''_{5}$ of Cr(II) spins, and below 291 K the $\Gamma''_{10}$ of Cr(II) spins.
The three decomposed IRs of Cr(I) spins with wave-vector $(\frac{1}{2}, \frac{1}{2}, 0)$ and four IRs of Cr(II) spins with wave-vector $(1, 0, 0)$ are illustrated in Table \ref{irrep_tb}.

The magnetic structure of Sr$_{2}$Cr$_{3}$As$_{2}$O$_{2}$ make interesting comparison to other `2322' compounds such as the manganese analog $A_2$Mn$_3Pn_2$O$_2$ and chromium pnictides LaCrAsO and BaCr$_{2}$As$_{2}$.
In $A_2$Mn$_3Pn_2$O$_2$ and chromium pnictides, the magnetic moments in the Mn$_2$As$_2$ and Cr$_2$As$_2$ sublattices form AFM arrangement both within and between layers with the wave-vector (1,0,1) in the so-called G-type magnetic structure \cite{Brock_magstr1996,PhysRevB.81.224513.2010,LnCrAsO.2013,PhysRevB.95.184414}.
In Sr$_{2}$Cr$_{3}$As$_{2}$O$_{2}$, however, the inter-layer spin arrangement is FM and the magnetic structure becomes a C-type with the magnetic wave-vector (1,0,0).
The magnetic CrO$_2$ buffer layer between the Cr$_2$As$_2$ layers might be responsible for the change of the coupling sign.
The ordering temperatures for the Mn$_2$As$_2$ sublattice in $A_2$Mn$_3Pn_2$O$_2$ are around 300 K. 
But for the Cr$_2$As$_2$ sublattice in chromium pnictides and Sr$_{2}$Cr$_{3}$As$_{2}$O$_{2}$, the ordering temperatures are close to 600 K, indicating much stronger interaction strength in the Cr$_2$As$_2$ sublattice than the Mn$_2$As$_2$ sublattice.

The MnO$_2$ layers in Sr$_2$Mn$_3$Sb$_2$O$_2$, Sr$_2$Mn$_3$As$_2$O$_2$ and Ba$_2$Zn$_2$MnAs$_2$O$_2$ establish magnetic order with wave-vector $(\frac{1}{2}, \frac{1}{2}, 0)$ \cite{Brock_magstr1996,Ozawa2001,PhysRevB.81.224513.2010}, the same as the CrO$_2$ layer in Sr$_{2}$Cr$_{3}$As$_{2}$O$_{2}$.
However, only in Sr$_2$Mn$_3$Sb$_2$O$_2$ the magnetic order is long-ranged and at a much reduced temperature of 65 K. 
In Sr$_2$Mn$_3$As$_2$O$_2$, Sr$_2$Zn$_2$MnAs$_2$O$_2$ and Ba$_2$Zn$_2$MnAs$_2$O$_2$, the magnetic order is short-ranged with a spin freezing behavior.
These contrast to the long-range order at 291 K for the CrO$_2$ layer spins in Sr$_{2}$Cr$_{3}$As$_{2}$O$_{2}$.
Again the magnetic interactions in the CrO$_2$ sublattice of Sr$_{2}$Cr$_{3}$As$_{2}$O$_{2}$ are much stronger than in the MnO$_2$ sublattice in $A_2$Mn$_3Pn_2$O$_2$.

The moments in the Cr$_2$As$_2$ layer are along the $c$ direction above 291 K, consistent with the orientation in the `1111' and `122' chromium pnictides.
This reflects the same easy axis due to the single-ion anisotropy of Cr$^{2+}$ in the Cr$_2$As$_2$ environment.
However, the magnetic ordering of the Cr(I) ions along the $c$ direction in the CrO$_2$ sublattice flips the Cr(II) ions in the Cr$_2$As$_2$ sublattice into the $ab$ plane, changing the easy axis in the Cr$_2$As$_2$ layer.
The magnetic structure where two sublattices have orthogonal moment directions was also observed in Sr$_2$Mn$_3$Sb$_2$O$_2$ \cite{Brock_magstr1996}, although in a reverse way: 
the moments in the Mn$_2$As$_2$ layer order along the $c$ direction, and in the MnO$_2$ layer in the $ab$ plane.
It was believed previously that in the `2322' compounds the two sublattices show independent magnetic behaviors and the interaction between the two sublattices was negligible \cite{Brock_magstr1996}.
The abrupt spin-flop transition found in our work on Sr$_{2}$Cr$_{3}$As$_{2}$O$_{2}$ (Fig.~\ref{op} (b)) provides evidence that the two magnetic sublattices are closely correlated.
The magnetic state of orthogonal moment directions in the two sublattices was not reproduced through simple bilayer classical spin model with experimentally achievable interactions strengths \cite{Enjalran2000}.
The nature of interaction and origin for the orthogonal spin orientation thus deserve more investigations.

While $A_2$Mn$_3Pn_2$O$_2$ are insulators, Sr$_{2}$Cr$_{3}$As$_{2}$O$_{2}$ and other chromium pnictides are metallic.
We noticed that the conduction electrons are mostly from the Cr$_2$As$_2$ layer, while electrons of the CrO$_2$ layer are largely localized in the first principle calculations\cite{PhysRevB.92.205107.2015}.
The saturation moment is $2.2 \mu_{B}$ per Cr(II) in the Cr$_2$As$_2$ layer and $3.1 \mu_{B}$ per Cr(I) in the CrO$_2$ layer.
There is a larger reduction of the saturated moment in the itinerant Cr$_2$As$_2$ layer than in the localized Cr$_2$As$_2$ layer from the $4 \mu_{B}$ per Cr$^{2+}$ ion in the $3d^4$ high spin state.
Therefore, it is tempted to state that the Cr$_2$As$_2$ layer contributes itinerant moments and the CrO$_2$ layer local moments, although it could be more complex due to correlations and interplay of itinerant and local electrons from two sublattices.

\section{Conclusion}

We have performed neutron powder diffraction experiments from 4 to 620 K to investigate magnetic structures and magnetic transitions in Sr$_{2}$Cr$_{3}$As$_{2}$O$_{2}$. 
We discovered successive magnetic transitions in the material. The Cr(II) ions in the Cr$_2$As$_2$ sublattice develop a C-type antiferromagnetic order of the magnetic wave-vector $(1,0,0)$ at 590.3(7) K with
the magnetic easy axis along the $c$ direction.
Below 291 K, a K$_2$NiF$_4$-like antiferromagnetic order of the Cr(I) ions occurs with the magnetic wave-vector $(\frac{1}{2}, \frac{1}{2}, 0)$ in the CrO$_2$ sublattice with the easy $c$-axis.
At the same transition at 291 K, magnetic moments of the Cr(II) ions in the Cr$_2$As$_2$ sublattice flip into the $ab$ plane.
The magnetic structure of orthogonal moment orientation in the two sublattices persists down to 4 K.
The interplay between the two magnetic sublattices may play an important role in facilitating the spin-flop transition of Cr(II) moments. 

%Theoretic works have not been able to correctly predict the magnetic structures and phase-transitions observed in this work. Further investigation is called for.
%the correlation effects and the role of itinerant and local moments

\begin{acknowledgments}

The work at RUC was supported by the National Basic Research Program of China (Grant No.~2012CB921700), the National Natural Science Foundation of China (Grant No.~11190024) and the Key Laboratory of Neutron Physics, CAEP (Grant No. 2017CB01). J.~L.~and J.~W.~acknowledges support from the Research Funds of RUC (Grant No.~17XNLF04 and 17XNLF06). J.~S.~acknowledges support from China Scholarship Council. The work at ZU was supported by the National Key Research and Development Program of China (No. 2017YFA0303002).
Research at Oak Ridge National Laboratory was sponsored by the Scientific User Facilities Division, Office of Basic Energy Sciences, U.S. Department of Energy.

\end{acknowledgments}


\end{document}